# Observation of Parity-Time Symmetry in Optically Induced Atomic Lattices

Zhaoyang Zhang,[1,2] Yiqi Zhang,[2] Jiteng Sheng,[3] Liu Yang,[1,4] Mohammad-Ali Miri,[5] Demetrios N. Christodoulides,[5] Bing He,[1] Yanpeng Zhang,[2] and Min Xiao[1,6,*]

[1]*Department of Physics, University of Arkansas, Fayetteville, Arkansas 72701, USA*
[2]*Key Laboratory for Physical Electronics and Devices of the Ministry of Education & Shaanxi Key Lab of Information Photonic Technique, Xi'an Jiaotong University, Xi'an 710049, China*
[3]*Department of Physics and Astronomy, The University of Oklahoma, Norman, Oklahoma 73019, USA*
[4]*College of Physics, Jilin University, Changchun 130012, China*
[5]*CREOL, College of Optics and Photonics, University of Central Florida, Orlando, Florida 32816, USA*
[6]*National Laboratory of Solid State Microstructures and School of Physics, Nanjing University, Nanjing 210093, China*

We experimentally demonstrate $PT$-symmetric optical lattices with periodical gain and loss profiles in a coherently prepared four-level $N$-type atomic system. By appropriately tuning the pertinent atomic parameters, the onset of $PT$-symmetry breaking is observed through measuring an abrupt phase-shift jump between adjacent gain and loss waveguides. The experimental realization of such a readily reconfigurable and effectively controllable $PT$-symmetric waveguide array structure sets a new stage for further exploiting and better understanding the peculiar physical properties of these non-Hermitian systems in atomic settings.

The discovery of parity-time ($PT$) symmetric Hamiltonians has allowed the physics community to see the behaviors of non-Hermitian systems in a new light [1]. It was found that a broad class of complex Hamiltonians can exhibit altogether real eigenvalue spectra, provided they respect the parity-time symmetry [1–3]. Another intriguing consequence of such a $PT$-symmetry condition is the possibility for an abrupt symmetry breaking phase transition, beyond which the spectrum ceases to be real and starts to become complex, once a parameter controlling the degree of non-Hermiticity exceeds a certain critical threshold [4–6]. The experimental studies came about only recently after recognizing that optics can provide a fertile ground where $PT$-symmetric concepts can be implemented [4–8]. What facilitates this possibility is the formal equivalence between the quantum Schrödinger equation and the optical wave propagation equation (under paraxial approximation), where the gain-loss parameters are responsible for introducing the non-Hermiticity into the systems. Based on this isomorphism, one can easily show that a necessary (yet not sufficient) condition for $PT$ symmetry is that the real part of the complex potential must be an even function of the position while its imaginary counterpart (in optics corresponding to the gain or absorption) must have a spatially antisymmetric profile [3,5]. In recent years, significant progress has been made on both theoretical and experimental fronts concerning a variety of optical $PT$-symmetric systems that simultaneously engage gain and loss processes in a balanced fashion [7–24]. These studies have unveiled a number of interesting phenomena such as non-Hermitian Bloch oscillations [7,8], unidirectional invisibility [14–17], perfect laser absorbers [20,21], optical solitons [22], and non-Hermitian manifestations of topological insulators [23] in $PT$-symmetric optical configurations. More recently, an "exceptional ring" effect was reported in a Dirac cone setting analogous to the gain-loss structure in $PT$-symmetric optics [24].

Since the refractive index, particularly the gain or loss properties, can be simultaneously manipulated in multilevel atomic systems [25,26], realizations of $PT$-symmetric potentials [with $n(x) = n^*(-x)$] have been theoretically proposed in certain multilevel atomic configurations [27–30]. Quite recently, an anti-$PT$-symmetric potential [with $n(x) = -n^*(-x)$] has been experimentally produced in a pair of optically induced waveguides coupled by flying atoms [31]. Compared with solid-state systems, $PT$ symmetry in atomic media can possess certain distinguished features due to their intrinsic attributes, such as the light-induced atomic coherence, which can result in easily controllable absorption, dispersion, Raman gain, and nonlinearity. First, we are able to construct gain- and loss-modulated optical lattices in an $N$-type atomic system and demonstrate the true $PT$ symmetry and its breaking in a simultaneous gain and loss optical waveguide array, which was extensively studied theoretically [32,33] but has not yet been experimentally achieved. Second, with multiple tunable parameters, the atomic system allows real-time reconfigurable capability and easy tunability (especially for the periodicity and structure of the lattice) without employing sophisticated fabrication technologies and making a large number of samples, which provide a new platform to study $PT$ symmetry under different parametric regimes and other non-Hermitian Hamiltonians. Third, many interesting effects, such as nonlinear $PT$-symmetric defect modes [34], solitons in $PT$-symmetric nonlinear lattices [35], and unidirectional light transport [36], have been predicted recently by considering the interplays between *PT*-symmetric potential and Kerr nonlinearity. Such phenomena might be relatively easy to be observed in electromagnetically induced transparency [37] (EIT) atomic systems with enhanced and controllable nonlinearity [38], which opens the door for future experimental studies of non-Hermitian nonlinear optics.

Inspired by recent developments in *PT*-symmetric optics and the superiorities of atomic media, here, we experimentally demonstrate light wave transport in a periodic *PT*-symmetric potential in a four-level atomic configuration driven by a weak signal field and two sets of standing-wave (coupling and pump) laser fields [39]. The *PT*-symmetric potential is achieved by spatially engineering the desired complex refractive indices of the atomic assemble. The standing-wave coupling field propagating along the $z$ direction is responsible for establishing the optically induced lattice along the transverse direction $x$. By launching the weak signal field into the lattice, we can obtain discrete diffraction patterns [40], as well as the underlying spatially modulated susceptibility, under the EIT condition. Our results clearly indicate that by adding another standing-wave pump field, spatially periodic gain and loss regions with high contrast can be generated on the launched signal field. The induced spatially periodic *PT*-symmetric optical potential (with a periodic even refractive index and odd gain or loss profiles) can be produced by properly tuning the pertinent experimental parameters. The manifestation of the spontaneous *PT*-symmetry breaking phenomenon is directly observed by monitoring the relative phase difference between the adjacent gain and loss channels. This is accomplished by interfering the signal beam passing through the atomic medium with a reference beam in the $y$ direction. The experimental observations can be well explained through numerical simulations.

Figure 1(a) schematically depicts the experimental setup. The signal field and two sets of standing-wave fields propagating along the same $z$ direction interact with an $N$-type four-level $^{85}$Rb atomic system [see Fig. 1(b)], which consists of two hyperfine states $F = 2$ (level $|1\rangle$) and $F = 3$ ($|2\rangle$) of the ground state $5S_{1/2}$ and two excited states $5P_{1/2}$ ($|3\rangle$) and $5P_{3/2}$ ($|4\rangle$). Two elliptical-Gaussian-shaped coupling beams $E_c$ and $E'_c$ (of wavelength $\lambda_c = 794.97$ nm, frequency $\omega_c$, and Rabi frequencies $\Omega_c$ and $\Omega'_c$, respectively) from the same external cavity diode laser are symmetrically placed with respect to the $z$ axis and intersect at the center of the rubidium cell at an angle of $2\theta \approx 0.4°$ to establish an optical lattice along the transverse direction $x$ inside the cell. Similarly, two pump beams $E_p$ and $E'_p$ ($\lambda_p = 780.24$ nm, $\omega_p$, and $\Omega_p$ and $\Omega'_p$), partially overlapped with $E_c$ and $E'_c$, respectively, enter the cell at almost the same angle $2\theta$ to form a pump-field optical lattice. The 7 cm long atomic vapor cell wrapped with $\mu$-metal sheets is heated by a heat tape to provide an atomic density of $\sim 2.0 \times 10^{12}$ cm$^{-3}$ at 75 °C. The signal beam $E_s$ ($\lambda_s = 794.97$ nm, $\omega_s$, $\Omega_s$) with a Gaussian intensity profile

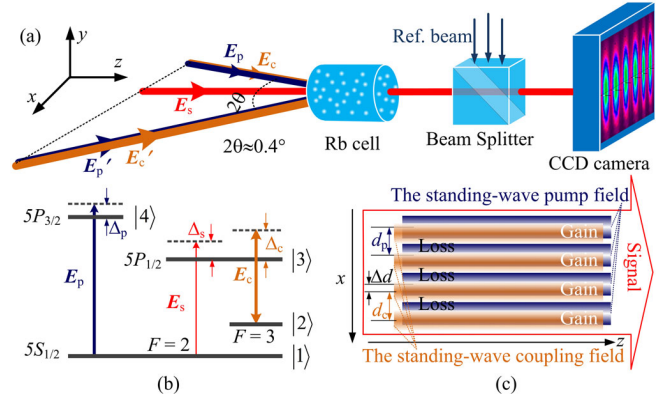

FIG. 1. (a) Experimental setup. $E'_p$ and $E_p$ are pump beams from the same laser, and so are the coupling beams $E'_c$ and $E_c$. The reference beam intersects with $E_s$ to generate the reference interference. (b) The energy-level diagram of the four-level $N$-type configuration in $^{85}$Rb atomic vapor. (c) The spatial arrangements of the signal field, coupling-field lattice, and pump-field lattice. $x$ and $z$ represent the transverse and longitudinal directions of beam propagation, respectively. $\Delta d$ marks the displacement between the two lattices.

propagates through the two sets of optical lattices, as shown in Fig. 1(c).

By properly adjusting the experimental parameters, active Raman gain, one of the most important requirements for implementing the exact *PT* symmetry in optics, can be generated on the signal field [26,39]. As a result, the desired periodic gain and loss profiles along the $x$ direction are obtained after $E_s$ passes through these two partially overlapping optical lattices. The periodically gain- and loss-modulated $E_s$ then interferes with a reference beam (injected in the $y$ direction) to exhibit the induced phase difference between the adjacent gain and loss channels. The reference beam originates from the same external cavity diode laser as $E_s$ and is introduced into the optical path via a 50/50 beam splitter to intersect with $E_s$ at the position of a charge coupled device (CCD) camera [see Fig. 1(a)], which is used to monitor both the output signal beam and the relative phase difference. Figure 1(c) shows a schematic diagram of the spatial arrangement for the two sets of optical lattices and $E_s$ inside the cell. The spatial periodicity of the coupling lattices is $d_c = \lambda_c/2 \sin\theta \approx 114$ $\mu$m, and the spatial-shift distance $\Delta d$ between the two lattices can be adjusted to control the real and imaginary parts of the susceptibility experienced by $E_s$.

Figures 2(a) and 2(b) show the calculated real and imaginary parts of the susceptibility versus the signal-field frequency detuning for different pump-beam intensities. The frequency detunings for the signal, coupling, and pump fields are defined as $\Delta_s = \omega_{31} - \omega_s$, $\Delta_c = \omega_{32} - \omega_c$, and $\Delta_p = \omega_{41} - \omega_p$, respectively. To achieve the *PT*-symmetric conditions in the current atomic lattices, the values of the real part versus $\Delta_s$ at $\Omega_p = 0$ and $\Omega_p \neq 0$ must be the same, while the corresponding imaginary parts must have the same absolute value but opposite sign. The theoretically calculated spatial distributions of the refractive index in Figs. 2(c) and 2(d) clearly indicate that a *PT*-symmetric structure with alternating gain and loss waveguides [39] can indeed be established in such an atomic configuration. The periodic structure can be interpreted through a complex potential $V(x)$ in the paraxial wave equation (5), which is mathematically isomorphic to the Schrödinger equation:

$$i\frac{\partial E}{\partial z} + \frac{\partial^2 E}{\partial x^2} + V(x)E = 0. \quad (1)$$

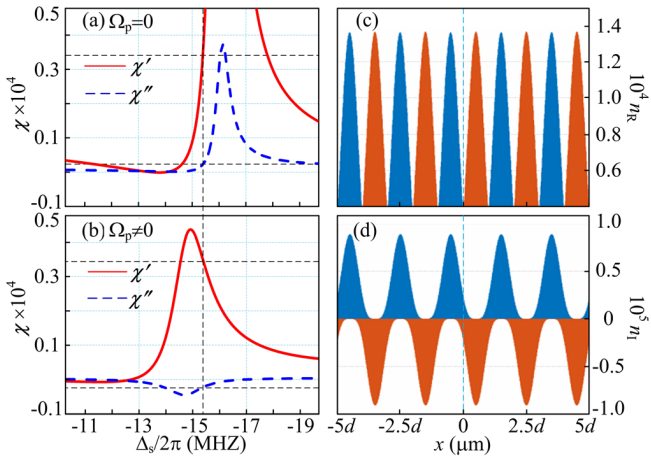
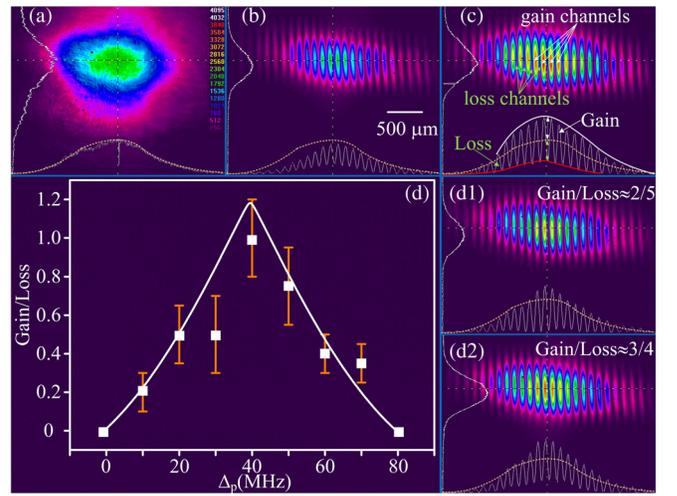

FIG. 2. The real (dispersion) and imaginary (gain or absorption) parts of the susceptibility versus $\Delta_s$ at (a) $\Omega_p = 0$ and (b) $\Omega_p \neq 0$, respectively. The (c) real and (d) imaginary parts for periodic lattices of the refractive index as a function of position $x$ with the coupling intensities spatially modified. $\Omega_s = 2\pi \times 10$ MHz, $\Omega_c = 2\pi \times 0.2[1 + \cos(\pi x/2)]$ MHz, $\Omega_p/2\pi = 6$ MHz, $\Delta_p = 40$ MHz, $\Delta_c = -100$ MHz, and $\Delta_s \approx -2\pi \times 15$ MHz.

FIG. 3. (a) Image and intensity profile of $E_s$ without interacting with atoms. (b) Signal beam after propagating through the coupling lattice. (c) Simultaneous gain and loss profiles on $E_s$ with both lattices turned on. (d) Dependence of the gain-to-loss ratio on $\Delta_p$. The squares are experimental observations, and the solid curve is the theoretical prediction. The observed gain-loss profiles are presented at (d1) $\Delta_p = 30$ MHz and (d2) 10 MHz.

Here, the electric field envelope can be written as

$$E(x,z,t) = \exp(i\beta z)\sum_{j=1}^{10} A_j(x)E_j(z), \quad (2)$$

where $E_j(x)$ represents the eigenmode field profile in each waveguide element and $A_j(z)$ denotes the corresponding modal amplitude in this channel. Based on Eqs. (1) and (2), one can then write down a coupled-mode (tight-binding) set of equations from which the complex band structure can be predicted. For a ten-waveguide coupled system (which can be readily implemented in our experiment), an exceptional point (where the *PT* symmetry breaks) exists at $\gamma/2\kappa = 0.284$, where $\gamma = \gamma_G = \gamma_L$ is the gain or loss coefficient and $\kappa$ is the coupling coefficient between the adjacent waveguides. In principle, an infinite number of coupled waveguides can be considered. However, due to the limited beam sizes and the periodicity of the waveguides, and considering the alternate gain-loss requirement for the *PT* symmetry, we use ten effective waveguides in the theoretical model to mimic the experiment. Numerical simulations indicate that the breaking threshold decreases as the number of waveguides increases (see details in the Supplemental Material [41]).

In the experiment, we first set a periodic refractive index modulation based on the EIT scheme (generated by the signal and coupling fields) [43] and then establish the periodic gain-loss profiles (by adding the standing-wave pump field), both along the $x$ direction. With the signal beam [as shown in Fig. 3(a)] first launched into the coupling lattice (with the pump fields blocked), we observed the discrete diffraction pattern that manifests the periodic modification of the signal-field refractive index. Such discrete diffraction patterns appear within a frequency detuning window of about 50 MHz near the two-photon resonance satisfying $\Delta_s - \Delta_c = 0$ [40]. With the signal-field detuning set as $\Delta_s = -100$ MHz, the diffraction image shown in Fig. 3(b) is obtained by carefully adjusting $d_c$ to match the maximum refractive index contrast at $\Delta_s - \Delta_c = 10$ MHz.

The presence of the pump-field lattice can provide an amplification for $E_s$. With the two sets of lattices turned on concurrently, we can simultaneously induce gain and loss regions with a high and controllable contrast on $E_s$ by carefully modifying the displacement $\Delta d$ between the two established optical lattices [see Fig. 1(c)] and other experimental controlling parameters. As shown in Fig. 3(c), two adjacent channels in the lattice array experience alternative gain and loss, which can be determined by comparing the intensity profile of the signal beam before its interaction with the medium [Fig. 3(a)]. Figure 3(d) demonstrates the evolution of the gain-to-loss ratio as a function of $\Delta_p$—showing a sensitive dependence. Figures 3(d1) and 3(d2) present the observed gain and loss intensity profiles at $\Delta_p = 30$ and 10 MHz, respectively. The gain-to-loss ratio in Fig. 3(c) reaches near unity, i.e., a balanced gain and loss between neighboring waveguides, which is required for achieving an exact $PT$ symmetry in such a coupled-waveguide array system [5].

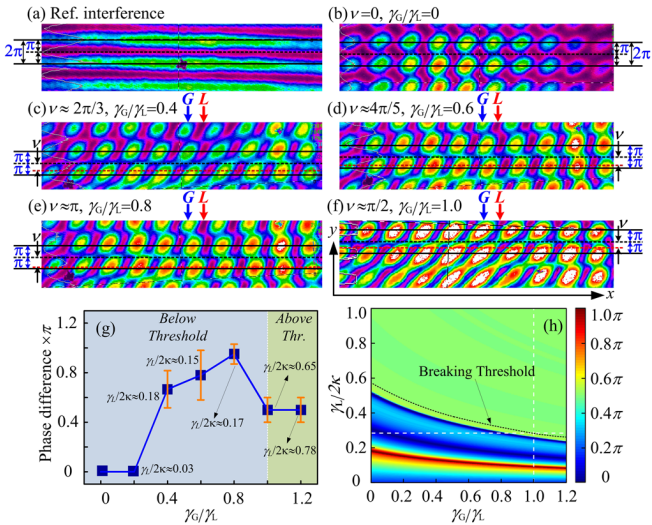

FIG. 4. Selected gain and loss channels for measurements are marked with $G$ and $L$, respectively. (a) Reference interference fringes generated by the reference beam and the signal beam without diffraction. The two solid lines mark the centers of two adjacent fringes. (b)-(f) Observed phase differences (marked by the pair of one-way arrows) between the adjacent gain and loss channels with $\gamma_G/\gamma_L$ being 0, 0.4, 0.6, 0.8, and 1.0, respectively. (g) Measured dependence of phase difference on $\gamma_G/\gamma_L$. The left (gray) and right (green) parts are the regions below and above the $PT$-symmetry breaking threshold, respectively. (h) Theoretical simulations of phase difference according to the coupled equations for ten waveguides at a propagation distance of $z = 10$. The breaking threshold (the dotted curve) decreases with an increasing gain-to-loss ratio. The dimensionless $z$ is scaled by the Rayleigh range $kx_0^2$ ($k = 2\pi/\lambda$, and $x_0$ is the waveguide width).

In a $PT$-symmetric system, the evolutions of eigenvalues can be the most reliable critera to determine whether the system is below or above the threshold. For such a non-Hermitian gain- and loss-modulated array, the behaviors of its eigenvectors can be captured by the changes in the relative phase difference $\nu$ (which represents the internal phase difference of the eigenvectors) between the adjacent gain and loss channels [12]. The distance (along the $y$ direction) between the adjacent interference fringes [as shown in Fig. 4(a)] is defined as $2\pi$. The dotted black line gives the center of the two solid lines, so the distance between the solid line and the dotted line is $\pi$. With the intensity of $E_s$ spatially modulated, the interference pattern divides the "bright" gain regions and "dark" loss regions into a netlike square lattice. A dotted red line is drawn along the center of a dark square in one row of the lattice to mark its position. The relative distance [referred to as the relative phase difference, marked by a pair of one-way arrows in Figs. 4(c)–4(e)] between the dotted red line and dotted black line represents the relative phase difference between two neighboring gain and loss channels [12].

Figure 4(b) shows the case that no phase difference is detected between the gain and loss channels when the gain is zero; i.e., the dotted black line and dotted red line overlap. Several relative phase differences ($\nu$) are measured in Figs. 4(c)–4(e) as the gain-to-loss ratio $\gamma_G/\gamma_L$ increases but still remains below the $PT$-symmetry breaking threshold. Figure 4(f) illustrates the situation above the breaking point, in which case the phase difference is fixed at $\nu = \pi/2$ even when the gain-to-loss ratio becomes slightly above unity (the measured $\nu$ stays unchanged as $\gamma_G/\gamma_L$ continuously increases from 1 to 1.2). The measured phase-difference dependence on $\gamma_G/\gamma_L$, as shown in Fig. 4(g), can be qualitatively explained by the theoretical predictions given in Fig. 4(h), illustrating that the value of $\gamma_L/2\kappa$ can indeed determine the relative phase difference produced in a coupled-waveguide array system with a certain $\gamma_G/\gamma_L$. The vertical axis $\gamma_L/2\kappa$ represents the evolution of $n_I/n_R$ since the coupling coefficient $\kappa$ directly relates to $n_R$. Since $\nu$ is consistently 0 under low and no gain conditions, there are no error bars for the first two data points.

For the point at $\gamma_G = \gamma_L$, the phase difference can vary from 0 to $\pi$ below the $PT$-symmetry breaking threshold and jumps to a fixed value of $\pi/2$ above threshold by increasing $\gamma_L/2\kappa$ [12]. Actually, Fig. 4(f) shows the exact $PT$-symmetry breaking point with $\gamma_G/\gamma_L = 1$ and $\gamma_L/2\kappa > 0.284$ simultaneously realized. For the cases of $\gamma_G \neq \gamma_L$, the coupled gain and loss waveguides can still have phase differences in the same way as the case of $\gamma_G = \gamma_L$. It is worth mentioning that the $PT$-symmetry breaking threshold value for can change with the value of the gain-to-loss ratio as indicated by the dotted curve shown in Fig. 4(h). Giving the experimental parameters at a certain gain-loss ratio in Fig. 4(g), we can calculate the $\gamma_L/2\kappa$ value and determine whether the system operates below or above the $PT$-symmetry breaking threshold according to the coupling equations [Eq. (S4) in the Supplemental Material]. In principle, even if a coupled-waveguide lattice system has an unbalanced gain-to-loss ratio (i.e., $\gamma_G/\gamma_L \neq 1$), one can still mathematically transform the system into a $PT$-symmetry-like configuration [12]. This then establishes a "quasi-$PT$-symmetry" system [44], in which the characteristic eigenvalue pattern is simply offset with respect to the original zero line [45]. Note that the dynamical behaviors of the exact $PT$-symmetry system and its quasi-$PT$-symmetry counterparts are essentially identical if the $PT$ symmetry is unbroken, while their dynamics are different when the $PT$ symmetry is broken [44].

In summary, we have experimentally demonstrated $PT$-symmetric optical lattices with a controllable gain-to-loss ratio in a coherently prepared $N$-type atomic ensemble. The required index modulation and the antisymmetric gain and loss profiles are introduced by exploiting the modified absorption (or EIT) and induced active Raman gain in the four-level atomic configuration. The presence of a well-defined breaking phase threshold was experimentally verified by observing the abrupt change of relative phase difference between the gain and loss channels. The constructed *PT*-symmetric atomic lattices can be used to study a variety of effects related to *PT* symmetry and other non-Hermitian Hamiltonians, including anti-*PT*-symmetric lattice and the *PT*-symmetric Talbot effect [46] as well as intriguing beam dynamical features [5] such as double refraction, power oscillation, and nonreciprocal diffraction patterns.

M. X. and Y. P. Z. acknowledge partial support from the NBRPC (No. 2012CB921804). M. X. was partially supported by the NSFC (No. 61435007 and No. 11321063). The work finished at Xi'an Jiaotong University was supported by KSTIT of Shaanxi Province (2014KCT-10) and the NSFC (No. 11474228, No. 61308015 and No. 61605154). D. N. C. and M.-A. M. were partially supported by the NSF (Grant No. ECCS-1128520) and the AFOSR (FA9550-14-1-0037).

Z. Y. Z. and Y. Q. Z. contributed equally to this work.

# Supplementary Information for "Observation of Parity-Time Symmetry in Optically Induced Atomic Lattices"

Zhaoyang Zhang[1, 2†], Yiqi Zhang[2†], Jiteng Sheng[3], Liu Yang[1, 4], Mohammad-Ali Miri[5], Demetrios N. Christodoulides[5], Bing He[1], Yanpeng Zhang[2] and Min Xiao[1, 6*]

[1]*Department of Physics, University of Arkansas, Fayetteville, Arkansas 72701, USA*
[2]*Key Laboratory for Physical Electronics and Devices of the Ministry of Education & Shaanxi Key Lab of Information Photonic Technique, Xi'an Jiaotong University, Xi'an 710049, China*
[3]*Department of Physics and Astronomy, The University of Oklahoma, Norman, OK 73019, USA*
[4] *College of Physics, Jilin University, Changchun 130012, China*
[5]*CREOL, College of Optics and Photonics, University of Central Florida, Orlando, Florida 32816, USA*
[6]*National Laboratory of Solid State Microstructures and School of Physics, Nanjing University, Nanjing 210093, China*
[†]Contributed equally to this work; [*]Corresponding author: *mxiao@uark.edu*

## 1. Theoretical derivations of the refractive index profile with PT-symmetric configuration in the four-level *N*-type atomic system

The density-matrix equations for the four-level *N*-type atomic system (see Fig. 1(b) in the manuscript) under the rotating-wave approximation are given by

$$\dot{\rho}_{22} = \Gamma_{42}\rho_{44} + \Gamma_{32}\rho_{33} - \Gamma_{21}\rho_{22} + \frac{i}{2}(\rho_{32} - \rho_{23})\Omega_c,$$

$$\dot{\rho}_{33} = \Gamma_{43}\rho_{44} - \Gamma_{32}\rho_{33} - \Gamma_{31}\rho_{33} + \frac{i}{2}[(\rho_{23} - \rho_{32})\Omega_c + (\rho_{13} - \rho_{31})\Omega_s],$$

$$\dot{\rho}_{44} = -(\Gamma_{43} + \Gamma_{42} + \Gamma_{41})\rho_{44} + \frac{i}{2}(\rho_{14} - \rho_{41})\Omega_p,$$

$$\dot{\rho}_{21} = -\tilde{\gamma}_{21}\rho_{21} + \frac{i}{2}(\rho_{31}\Omega_c - \rho_{24}\Omega_p - \rho_{23}\Omega_s),$$

$$\dot{\rho}_{31} = -\tilde{\gamma}_{31}\rho_{31} + \frac{i}{2}[\rho_{21}\Omega_c - \rho_{34}\Omega_p + (\rho_{11} - \rho_{33})\Omega_s], \quad (S1)$$

$$\dot{\rho}_{41} = -\tilde{\gamma}_{41}\rho_{41} + \frac{i}{2}[-\rho_{43}\Omega_s + (\rho_{11} - \rho_{44})\Omega_p],$$

$$\dot{\rho}_{32} = -\tilde{\gamma}_{32}\rho_{32} + \frac{i}{2}[\rho_{12}\Omega_s + (\rho_{22} - \rho_{33})\Omega_c],$$

$$\dot{\rho}_{42} = -\tilde{\gamma}_{42}\rho_{42} + \frac{i}{2}(\rho_{12}\Omega_p - \rho_{43}\Omega_c),$$

$$\dot{\rho}_{43} = -\tilde{\gamma}_{43}\rho_{43} + \frac{i}{2}(\rho_{13}\Omega_p - \rho_{42}\Omega_c - \rho_{41}\Omega_s),$$

$$\rho_{11} + \rho_{22} + \rho_{33} + \rho_{44} = 1.$$

Here, $\Omega_s = \mu_{13}E_s/\hbar$, $\Omega_c = \mu_{23}E_c/\hbar$ and $\Omega_p = \mu_{14}E_p/\hbar$ are the Rabi frequencies corresponding to the signal, coupling and pump fields, respectively, $\mu_{ij}$ is the dipole momentum between levels $|i\rangle$ and $|j\rangle$. $\Gamma_{ij}$ is the decaying rate between $|i\rangle$ and $|j\rangle$, $\gamma_{ij}=(\Gamma_i+\Gamma_j)/2$ is the decoherence rate. $\tilde{\gamma}_{21} = \gamma_{21} - i(\Delta_s - \Delta_c)$, $\tilde{\gamma}_{31} = \gamma_{31} - i\Delta_s$, $\tilde{\gamma}_{41} = \gamma_{41} - i\Delta_p$, $\tilde{\gamma}_{32} = \gamma_{32} - i\Delta_c$, $\tilde{\gamma}_{32} = \gamma_{32} - i\Delta_c$, $\tilde{\gamma}_{32} = \gamma_{32} - i\Delta_c$, $\tilde{\gamma}_{42} = \gamma_{42} - i(\Delta_c + \Delta_p - \Delta_s)$, $\tilde{\gamma}_{43} = \gamma_{43} - i(\Delta_p - \Delta_s)$. $\Delta_s=\omega_s-\omega_{31}$, $\Delta_c=\omega_c-\omega_{32}$ and $\Delta_p=\omega_p-\omega_{41}$ are defined as the frequency detunings of the signal, coupling and pump fields, respectively. According to the relation $2N\mu_{13}\rho_{31}=\varepsilon_0\chi E_s$, the corresponding susceptibility can be obtained by numerically solving $\rho_{31}$ in Eq. (S1) under steady-state approximation. The initially calculated susceptibility is shown in Fig. S1. By comparing Figs. S1(b) and 1(d), we can see that the presence of pump field can give rise to simultaneous gain and loss in the system, and the zero point of the imaginary part keeps constant at different $\Omega_c$.

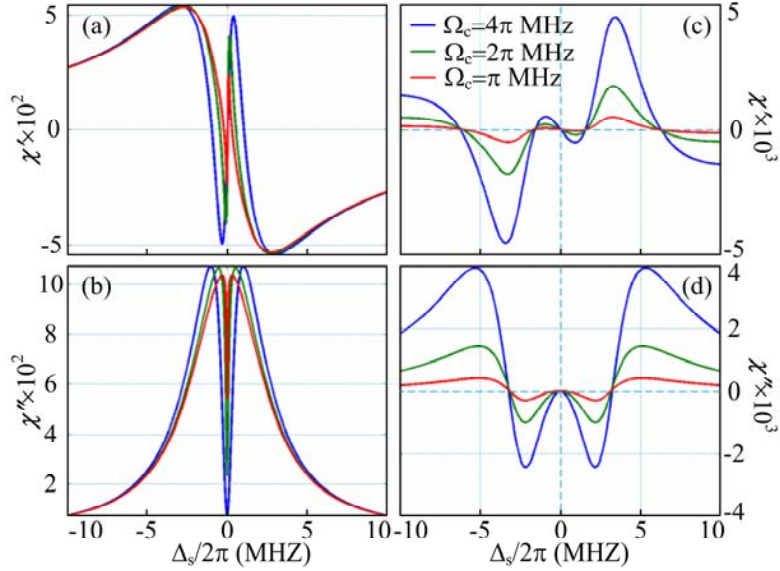

Figure S1. The theoretically calculated susceptibility $\chi$ versus $\Delta_s$. (a) Real part and (b) imaginary part of the susceptibility versus $\Delta_s$ with $\Omega_p=0$. (c) The real and (d) imaginary parts of the susceptibility versus $\Delta_s$ with $\Omega_p=2\pi\times6$ MHz. The presence of the pump field can result in positive and negative imaginary susceptibility at different $\Delta_s$ values. Other parameters are $\Omega_s=2\pi\times0.1$ MHz and $\Delta_p=\Delta_c=0$.

By properly setting the parameters as $\Delta_s\approx2\pi\times15$ MHz, $\Delta_c=-100$ MHz, $\Delta_p\approx40$ MHz, $\Omega_s=2\pi\times0.2[1+\cos(\pi x/2)]$ MHz and $\Omega_c=2\pi\times0.2$ MHz, the real and imaginary parts of the susceptibility can meet the condition for PT-symmetric potential, i.e. $\chi'$ has symmetric profile while the corresponding $\chi''$ becomes antisymmetric along the $x$ direction. The theoretical real and imaginary parts of the susceptibility corresponding to current experimental setup are shown as Fig. 2 in the manuscript.

## 2. Band structures of the periodical coupled gain-loss waveguide system

The above calculated complex spatial index refraction (susceptibility) represents the potential $V(x)$ in the paraxial wave equation, i.e. the Schrödinger-like equation [1-4]

$$i\frac{\partial E}{\partial z} + \frac{\partial^2 E}{\partial x^2} + V(x)E = 0. \tag{S2}$$

Considering the potential is uniform along the propagation direction, the *PT*-symmetric potential describes a periodic coupled-waveguide structure. In the potential, the electric field can be written as [1]

$$\begin{aligned}E(x,z,t) = \exp(i\beta z)[&A_1(x)E_1(z) + A_2(x)E_2(z) + A_3(x)E_3(z) + A_4(x)E_4(z) \\&+ A_5(x)E_5(z) + A_6(x)E_6(z) + A_7(x)E_7(z) + A_8(x)E_8(z) \\&+ A_9(x)E_9(z) + A_{10}(x)E_{10}(z),\end{aligned} \tag{S3}$$

where $A_m(x)$ is the eigenmode of each waveguide and $E_m(z)$ is the amplitude of the mode, $m=1, 2,\ldots,10$ is the number of the ten coupled waveguides. As a consequence, the coupling equations with 10 waveguides involved are given as [5, 6]

$$\begin{aligned}
i\frac{dE_1}{dz} - i\frac{\gamma_G}{2}E_1 + \kappa E_2 &= 0, \\
i\frac{dE_2}{dz} + i\frac{\gamma_L}{2}E_2 + \kappa(E_1 + E_3) &= 0, \\
i\frac{dE_3}{dz} - i\frac{\gamma_G}{2}E_3 + \kappa(E_2 + E_4) &= 0, \\
i\frac{dE_4}{dz} + i\frac{\gamma_L}{2}E_4 + \kappa(E_3 + E_5) &= 0, \\
i\frac{dE_5}{dz} - i\frac{\gamma_G}{2}E_5 + \kappa(E_4 + E_6) &= 0, \\
i\frac{dE_6}{dz} + i\frac{\gamma_L}{2}E_6 + \kappa(E_5 + E_7) &= 0, \\
i\frac{dE_7}{dz} - i\frac{\gamma_G}{2}E_7 + \kappa(E_6 + E_8) &= 0, \\
i\frac{dE_8}{dz} + i\frac{\gamma_L}{2}E_8 + \kappa(E_7 + E_9) &= 0, \\
i\frac{dE_9}{dz} - i\frac{\gamma_G}{2}E_9 + \kappa(E_8 + E_{10}) &= 0, \\
i\frac{dE_{10}}{dz} + i\frac{\gamma_L}{2}E_{10} + \kappa E_9 &= 0,
\end{aligned} \tag{S4}$$

where $\gamma_G$ and $\gamma_L$ are the gain and loss experienced by two adjacent waveguides (for example, the fifth ($A_5(x)$) and sixth ($A_6(x)$) waveguides) and $\kappa$ is the coupling coefficient. The three coefficients in Eq.(S4) explicitly are given as:

$$\gamma_G = \frac{\int V_6(x) A_5(x) A_6^*(-x)dx}{\int A_5(x) A_6^*(-x)dx},$$

$$\gamma_L = \frac{\int V_5(x) A_6(x) A_5^*(-x)dx}{\int A_6(x) A_5^*(-x)dx}, \quad (S5)$$

$$\kappa = \frac{\int V_5(x) A_6(x) A_6^*(-x)dx}{\int A_5(x) A_6^*(-x)dx}.$$

Also, we would like to note that the eigenmodes have the following relations:

$$A_1(x) = A_{10}^*(-x),\ A_2(x) = A_9^*(-x),$$
$$A_3(x) = A_8^*(-x),\ A_4(x) = A_7^*(-x), \quad (S6)$$
$$A_5(x) = A_6^*(-x),$$

and

$$A_1(x-4x_0) = A_3(x-2x_0) = A_5(x) = A_7(x+2x_0) = A_9(x+4x_0),$$
$$A_2(x-4x_0) = A_4(x-2x_0) = A_6(x) = A_8(x+2x_0) = A_{10}(x+4x_0). \quad (S7)$$

Here $x_0$ is the space between two adjacent waveguides.

According to the coupling equations in Eq. (S4), we can obtain the corresponding band structures (under balanced gain/loss case $\gamma=\gamma_G=\gamma_L$) shown in Figs. S2(e1) and S2(e2) in the manuscript, which clearly indicate that the exception point is at about $\gamma/2\kappa \approx 0.284$ when $N_{waveguide}=10$ waveguides are coupled. Also, according to Fig. S2, we can see that the exception point value decreases with increasing $N_{waveguide}$.

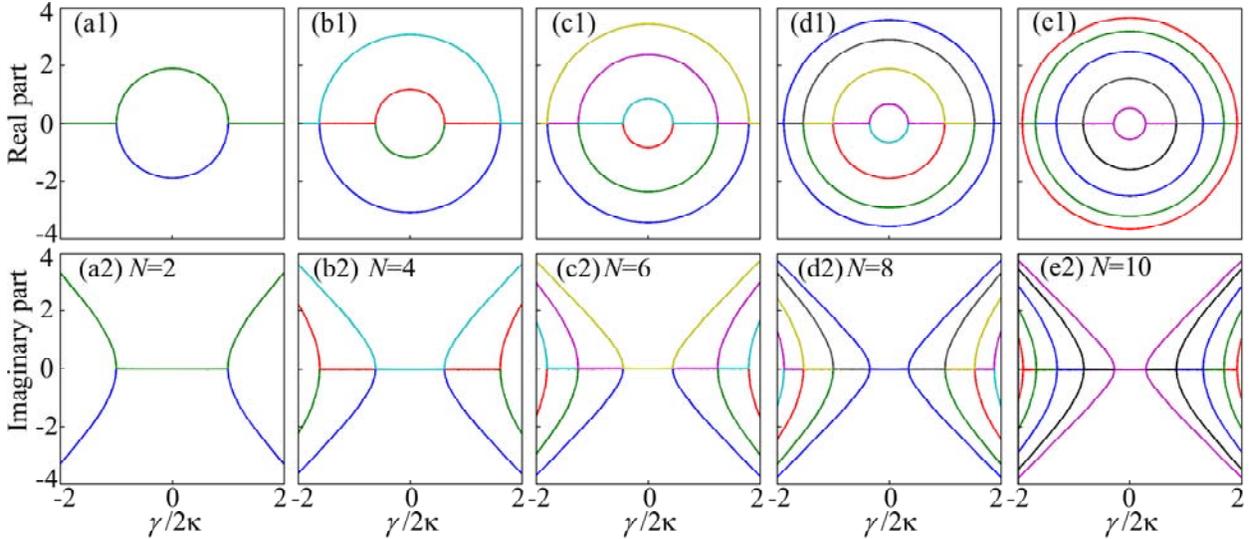

Figure S2. Band structures of the real and imaginary parts with $N_{waveguide}$ (=2, 4, 6, 8, 10) gain-loss waveguides coupled in an array. The PT-symmetry breaking threshold decreases as the number of waveguides increases.

### 3. Phase difference between the adjacent gain/loss waveguides

The phase difference (both below and above the PT-symmetry breaking threshold) between two neighboring waveguides is calculated according to the waveguides coupling equations in Eq. (S4). Figure S3 schematically shows how the phase difference is measured in experiment.

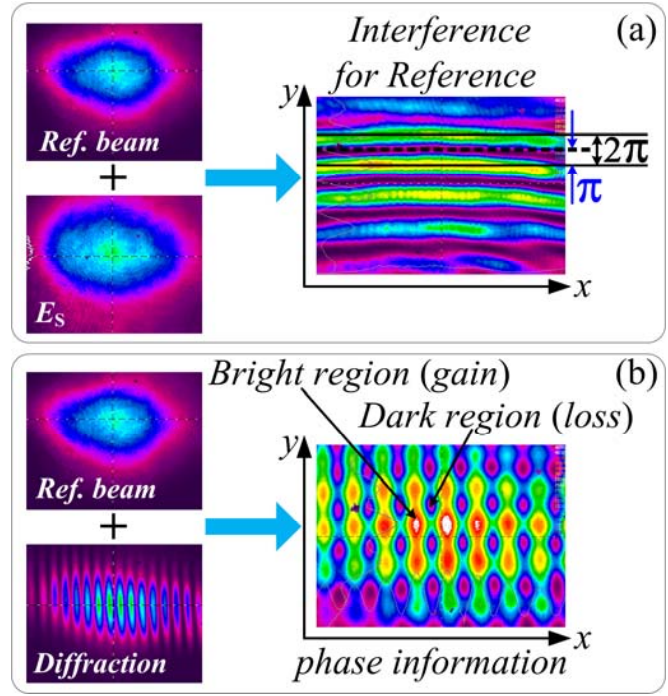

Figure S3. Schematic diagram for measuring the relative phase difference between the gain/loss channels. (a) The interference between the signal field $E_S$ and a reference beam (both of which are from the same laser) in the $y$ direction. The phase difference between the two solid lines is defined as $2\pi$. The phase difference between the black dotted line (located at the center of the two solid lines) and one of the two solid lines is $\pi$. (b) The interference pattern between the intensity modulated $E_S$ field (after diffraction) and the reference beam, so that the square-like lattice is obtained and the phase difference can then be measured.

### 4. Three sets of interference in the experiment.

There exist three interference patterns in the current experiment. First, the coupling beams $E_c$ and $E_c'$ (with vertical polarization) from the same continuous-wave diode laser (ECDL2) are coupled by two polarization beam splitters (PBSs) and intersect at the center of the vapor cell to establish the first interference pattern in the $x$ direction, namely, the standing-wave coupling field. The half-wave plates placed in front of the corresponding PBSs can adjust the powers of $E_c$ and $E_c'$. Second, the two pump beams $E_p$ and $E_p'$ (with horizontal polarization) from the same ECDL3 are coupled into the vapor cell by two reflective mirrors and build the standing-wave pump field with the powers of $E_p$ and $E_p'$ adjusted by rotating their corresponding half-wave plates. Third, we establish an interference pattern for reference outside the cell in the *y* direction by using the horizontally-polarized signal beam and the reference beam both from ECDL1.